\documentclass[longauth,traditabstract,letter]{aa}
\usepackage{longtable}
\usepackage{natbib,afterpage}
\usepackage{graphicx,color}
\usepackage{amssymb,times,graphics, subfigure}
\usepackage[english]{babel}
\usepackage{txfonts}

\def\Msun{\hbox{M$_{\odot}$}}               
\def\Rstar{\hbox{$R_{\star}$}}              
\def\Mdot{\hbox{$\dot{M}$}}               
\def\Teff{\hbox{$\rm{T}_{\rm eff}$}}            
\def\arcsec{\hbox{$^{\prime\prime}$}}

\bibpunct{(}{)}{;}{a}{}{,}

\begin{document}
\title{Water content and wind acceleration in the envelope around the oxygen-rich AGB star IK Tau as seen by Herschel/HIFI
\thanks{Herschel is an ESA space
  observatory with science instruments provided by European-led
  Principal Investigator consortia and with important participation
  from NASA.}}


   \author{
L.\ Decin \inst{1,2}
\and
K.\ Justtanont \inst{3}
\and
E. De Beck \inst{1}
\and
R. Lombaert \inst{1}
\and
A. de Koter \inst{2,4}
\and
L.B.F.M. Waters \inst{2}
\and
A.P. Marston \inst{5}
\and
D. Teyssier \inst{5}
\and
F.L. Sch\"oier
\inst{13}
\and
V. Bujarrabal    \inst{6}
\and
J. Alcolea \inst{7}
 \and
J. Cernicharo \inst{8}
\and
C. Dominik \inst{2,22}
\and
G. Melnick
\inst{9}
\and
K. Menten
\inst{10}
\and
D.A. Neufeld
\inst{11}
\and
H. Olofsson
\inst{12,13}
\and
P. Planesas
\inst{7,14}
     \and 
M. Schmidt
\inst{15}
     \and 
R. Szczerba \inst{15}
\and
 T. de Graauw   \inst{16} 
\and F. Helmich  \inst{17}              
\and P. Roelfsema     \inst{17}    
\and P. Dieleman  \inst{17}   
\and P. Morris  \inst{18}             
\and J.D. Gallego \inst{19}                
\and M.C. D\'{\i}ez-Gonz\'alez   \inst{19}    
\and E. Caux \inst{20,21}            
}


   \institute{   
Instituut voor Sterrenkunde,
             Katholieke Universiteit Leuven, Celestijnenlaan 200D, 3001
Leuven, Belgium \email{Leen.Decin@ster.kuleuven.be}
\and
    Sterrenkundig Instituut Anton Pannekoek, University of Amsterdam,
Science Park 904, NL-1098 Amsterdam, The Netherlands
\and
Onsala Space observatory, Chalmers University of Technology,
Dept. Radio \& Spece Science, S-439 92 Onsala, Sweden 
\and
Astronomical Institute, Utrecht University,
Princetonplein 5, 3584 CC Utrecht, The Netherlands 
\and
European Space Astronomy Centre, ESA, P.O. Box 78, E-28691
Villanueva de la Ca\~nada, Madrid, Spain
\and
Observatorio Astron\'omico Nacional. Ap 112, E-28803 
Alcal\'a de Henares, Spain\\
              \email{v.bujarrabal@oan.es}
	      \and
Observatorio Astron\'omico Nacional (IGN), Alfonso XII N$^{\circ}$3,
              E-28014 Madrid, Spain  
\and
Laboratory of Molecular Astrophysics, CAB, INTA-CSIC, 
Ctra de Torrej\'on a Ajalvir, km 4,
28850 Torrej\'on de Ardoz, Madrid, Spain
\and
Harvard-Smithsonian Center for Astrophysics, Cambridge, MA 02138, USA
\and
Max-Planck-Institut f{\"u}r Radioastronomie, Auf dem H{\"u}gel 69,
D-53121 Bonn, Germany 
\and
John Hopkins University, Baltimore, MD 21218, USA
\and
Department of Astronomy, AlbaNova University Center, Stockholm  
University, SE--10691 Stockholm, Sweden
\and
Onsala Space Observatory, Dept. of Radio and Space Science, Chalmers University of Technology, SE-43992 Onsala, Sweden
\and
Joint ALMA Observatory, El Golf 40, Las Condes, Santiago, Chile
\and
N. Copernicus Astronomical Center, Rabia{\'n}ska 8, 87-100 Toru{\'n}, Poland
\and
Atacama Large Millimeter/Submillimeter Array, 
Joint 
ALMA Office, Santiago, Chile
\and
SRON Netherlands Institute for Space Research, 
Landleven 
12, 9747 AD Groningen, the Netherlands
\and
Infrared Processing and Analysis Center, 
California 
Institute of Technology, MS 100-22, Pasadena, CA 91125,  USA
\and
Observatorio Astron\'omico Nacional (IGN), 
Centro 
Astron\'omico de Yebes, Apartado 148. 19080 Guadalajara, Spain
\and
Centre d'Etude Spatiale des Rayonnements, Universit\'e de Toulouse [UPS], 31062 Toulouse Cedex 9, France
\and 
CNRS/INSU, UMR 5187, 9 avenue du Colonel Roche, 31028 Toulouse Cedex 4, France
\and
Department of Astrophysics/IMAPP, Radboud University Nijmegen, P.O. Box 9010, 6500 GL Nijmegen, the Netherlands}

   \date{Received ... ; accepted ...}

 
  \abstract
{During their asymptotic giant branch, evolution low-mass stars lose a significant fraction of their mass through an intense wind, enriching the interstellar medium with products of nucleosynthesis. We observed the nearby oxygen-rich asymptotic giant branch star IK Tau using the high-resolution HIFI spectrometer onboard Herschel. We report  on the first detection of H$_2^{16}$O and the rarer isotopologues H$_2^{17}$O and H$_2^{18}$O in both the ortho and para states. We deduce a total water content (relative to  molecular hydrogen) of $6.6\times10^{-5}$, and an ortho-to-para ratio of 3:1. These results are consistent with the formation of H$_2$O in thermodynamical chemical equilibrium at photospheric temperatures, and does not require pulsationally induced non-equilibrium chemistry, vaporization of icy bodies or grain surface reactions. High-excitation lines of $^{12}$CO, $^{13}$CO, $^{28}$SiO,  $^{29}$SiO, $^{30}$SiO, HCN, and SO have also been detected. From the observed line widths, the acceleration region in the inner wind zone can be characterized, and we show that the wind acceleration is slower than hitherto anticipated.}

{}
{}
{}
{}

   \keywords{Line: profiles, Radiative transfer, Stars: AGB and post-AGB,
  (Stars): circumstellar matter, Stars: mass loss, Stars: individual:
  \object{IK Tau}}

	\authorrunning{Decin et al.}
	\titlerunning{Water content and wind acceleration in IK Tau}
   \maketitle
%

\section{Introduction}
IK Tau, also known as NML Tau, is an extremely red, oxygen-rich, Mira-type variable, with a period of about 470 days \citep{Wing1973ApJ...184..873W}. Its dust-driven wind produces a cool circumstellar envelope (CSE), which fosters a rich gas-phase chemistry \citep[e.g.,][]{Duari1999AandA...341L..47D}. IK Tau is relatively nearby, at a distance of $\sim$265\,pc \citep{Hale1997ApJ...490..407H}. Estimates of its mass-loss rates range from $3.8 \times 10^{-6}$ \citep{Neri1998AandAS..130....1N}  to $3 \times 10^{-5}$\,\Msun/yr \citep{GonzalesDelgado2003AandA...411..123G}. IK Tau's proximity and relatively high mass-loss rate facilitate the observation of molecular emission lines. Currently, a dozen different molecules and some of their isotopologs have been discovered in IK Tau, including CO, HCN, SiO, SiS, SO, SO$_2$, and NaCl \citep[e.g.][]{Milam2007ApJ...668L.131M,Decin2010}.

In this Letter, we report on the detection of thermal emission of water (H$_2$O) in the envelope around IK~Tau. The main isotopolog (H$_2^{16}$O) as well as the rarer isotopologs (H$_2^{17}$O and H$_2^{18}$O) are detected for both the ortho- and para-states.  We also present observations of high-excitation rotational transitions of $^{12}$CO, $^{13}$CO, $^{28}$SiO, $^{29}$SiO, $^{30}$SiO, HCN, and SO and demonstrate that the observed line widths characterize the acceleration region in the inner wind zone.

\section{Observations and data reduction} \label{observations}
The HIFI instrument \citep{DeGraauw2010} onboard the Herschel satellite \citep{Pilbratt2010} offers the possibility to observe molecular fingerprints  in the frequency ranges of 480--1150\,GHz and 1410--1910\,GHz at a spectral resolution up to 125\,kHz.
Single-point observations  towards IK\,Tau were carried out with the Herschel/HIFI instrument  in the dual beam switch (DBS) mode with a $3^{\prime}$ chop throw. The observation strategy and data-reduction are discussed in the online Appendix A and in \citet{Bujarrabal2010}.

\section{Results} \label{results}
In the 13 single-point observations obtained so far, covering a frequency range of 92.8 GHz in total, 31 molecular emission lines have been detected belonging to $^{12}$CO, $^{13}$CO, H$_2^{16}$O, H$_2^{17}$O, H$_2^{18}$O, $^{28}$SiO, $^{29}$SiO, $^{30}$SiO, HCN, SO, and NH$_3$ (see Table~\ref{Table:2} in the online Appendix). The detection of NH$_3$ is described in \citet{Menten2010}.

\begin{figure}
 \includegraphics[height=.48\textwidth,angle=90]{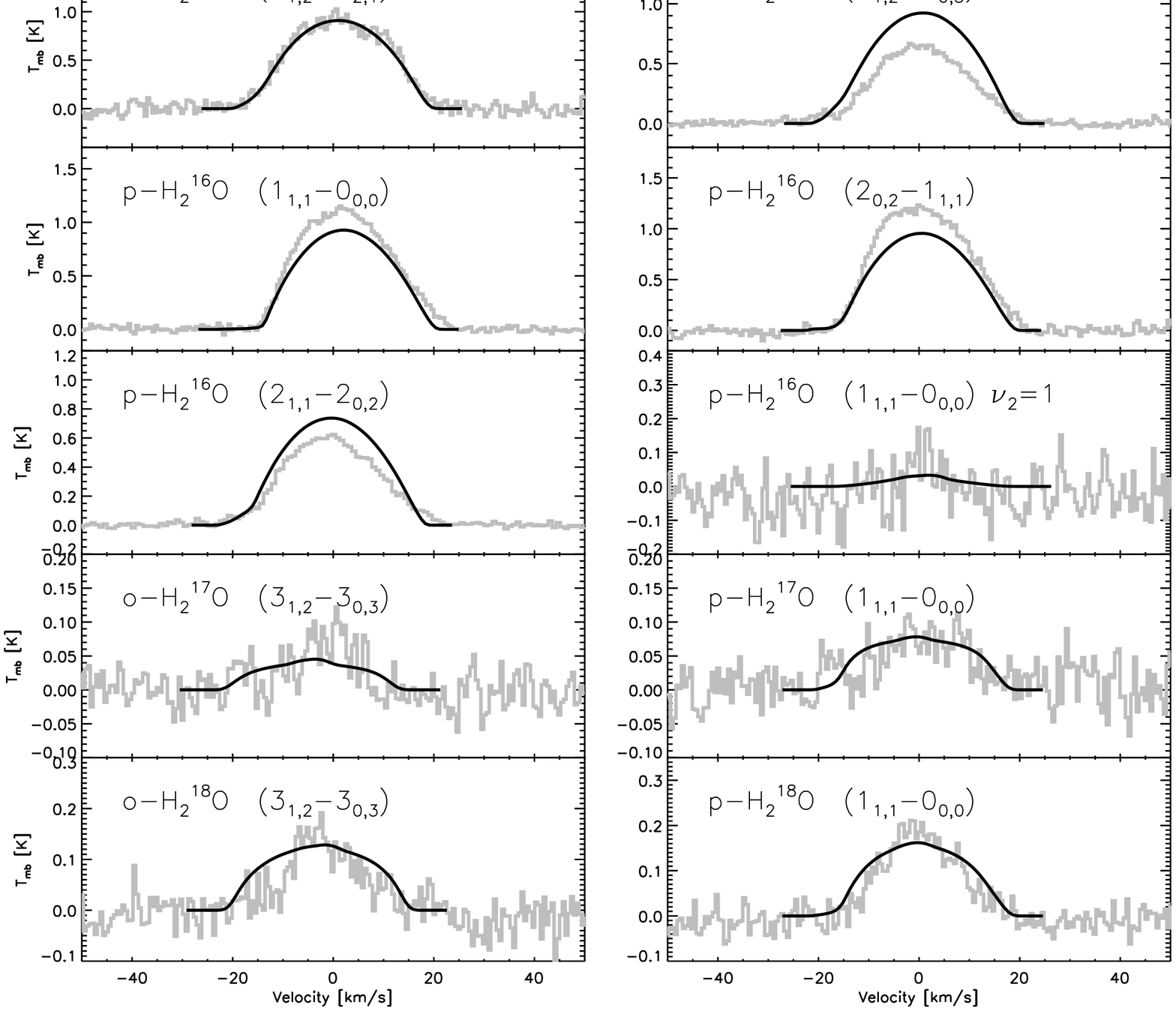}
 \vspace*{2ex}
\caption{HIFI observations of water lines (gray) compared to their  line profile predictions (black).
\emph{Row 1--2:} ortho-H$_2^{16}$O, \emph{row 3--4:} para-H$_2^{16}$O,  \emph{row 5:} ortho and para-H$_2^{17}$O, and \emph{row 6:} ortho and para- H$_2^{18}$O}
\label{Fig:1}
\end{figure}

For the first time, different excitation lines of water are discovered for the two nuclear spin isomers, ortho- and para-H$_2^{16}$O, as well as for the rare isotopologs H$_2^{17}$O and H$_2^{18}$O (see Fig.~\ref{Fig:1}).  The observations of these transitions provide information about the total water content, the ortho-to-para ratio,  and the isotopic ratios H$_2^{16}$O/H$_2^{17}$O and H$_2^{16}$O/H$_2^{18}$O (see Sect.~\ref{water}). 

\begin{figure*}
 \includegraphics[height=.96\textwidth,angle=90]{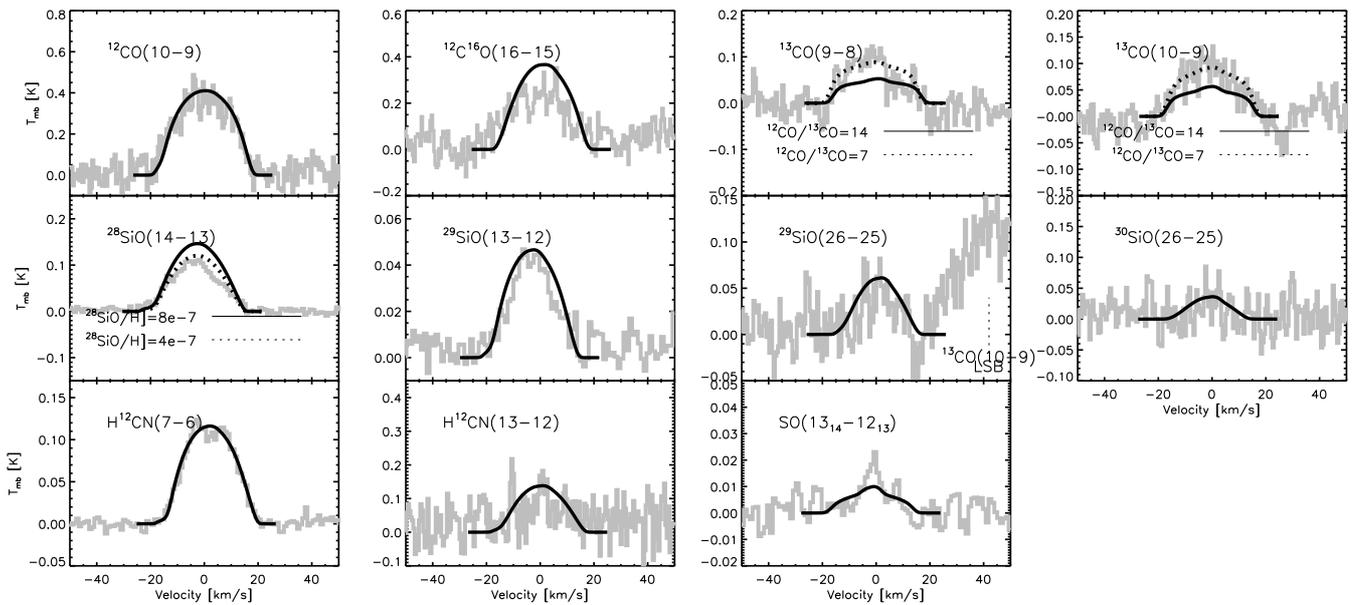}
\vspace*{2ex}
\caption{HIFI observations of line transitions of several molecules (gray) compared to their line profile predictions (black). Full black lines refer to line predictions using the molecular abundance stratifications as determined by \citet{Decin2010}, dotted black lines show the model predictions for the refined fractional abundances as determined in this Letter (see  the online Appendix B).}
\label{Fig:2}
\end{figure*}

High-excitation rotational transitions are observed for different molecules (see Fig.~\ref{Fig:2}). From the observed line widths, which range between 11 and 19\,km/s, it is immediately clear that the HIFI observations offer us a strong diagnostic to trace the wind acceleration zone in the inner envelope (see Sect.~\ref{envelope_structure}). Moreover, the high-excitation $^{12}$CO J=10--9 and J=16--15 lines, complemented with ground-based observations of  $^{12}$CO for J=1--0 to J=7--6, can be used as temperature indicator for the envelope as close as $\sim$20\,\Rstar\ (see Sect.~\ref{envelope_structure}).

\vspace*{-1.2cm}
\subsection{Water in IK Tau's envelope} \label{water}

\paragraph{Water content:} We observed both ortho and para-water lines for different isotopologs in IK Tau (Fig.~\ref{Fig:1}). Using the analytical formula from \citet{Groenewegen1994A&A...290..531G},  we calculated a photodissociation radius of $2.8 \times 10^{16}$\,cm or 1870\,\Rstar, while the theoretical models by \citet{Willacy1997AandA...324..237W} predict a value around 1600\,\Rstar. 
Assuming the same photodissociation radius for all isotopologs and isomers, we deduced (see Sect.~\ref{modelling} in the Appendix) a water-abundance [ortho-H$_2^{16}$O/H$_2$]=$5\times10^{-5}$ and an ortho-to-para water ratio (OPR) of 3:1. Taking the uncertainty in the photodissociation radius into account, the estimated uncertainty in the water abundance is a factor 2 for the assumed envelope structure (see Sect.~\ref{envelope_structure}).

The total (ortho+para) H$_2^{16}$O abundance for IK Tau derived in this Letter, relative to the total hydrogen content and assuming that all hydrogen is in the form of molecular hydrogen,   is [H$_2^{16}$O/H$_{\rm{tot}}$]=$3.3\times10^{-5}$, slightly lower than   the photospheric thermodynamic equilibrium (TE) prediction of $7\times10^{-5}$ for an evolved star with a C/O ratio of 0.75 \citep{Cherchneff2006AandA...456.1001C}. The theoretical TE value for the total water abundance is, however, very dependent on the exact value of the C/O ratio, which is unknown for IK Tau. \citet{Cherchneff2006AandA...456.1001C} predicts a higher inner wind abundance of $3.5\times10^{-4}$ around 5\,\Rstar\ in the case of pulsationally induced non-equilibrium chemistry. The upper limit implied by the cosmic abundances of carbon and oxygen is no more than $1 \times 10^{-3}$, assuming that all carbon is locked in CO and the remaining oxygen goes into H$_2$O \citep{Anders1989GeCoA..53..197A}. Vaporization of icy bodies \citep{Justtanont2005A&A...439..627J} or grain surface reactions via Fischer-Tropsch catalysis \citep{Willacy2004ApJ...600L..87W} will also result in a higher water abundance. Considering the agreement between the derived and the TE abundance, the effects of the other processes seem negligible for IK Tau.

On the basis of ISO-LWS data, \citet{Maercker2008A&A...479..779M} derived for IK Tau an ortho-H$_2^{16}$O abundance (relative to H$_2$) of $3.5 \times 10^{-4}$, significantly higher than the value we derived. Both studies used an almost equal photodissocation radius. However, \citet{Maercker2008A&A...479..779M}  only included the ground-state for the excitation analysis. As shown by \citet{Decin2010}, neglecting the first vibrational state of the bending mode, $\nu_2=1$, leads to an underprediction of the HIFI transitions here presented. Since the ISO data used in the study of \citeauthor{Maercker2008A&A...479..779M} are saturated, the integrated intensity is  insensitive to the exact value of the abundance, which will result in an overprediction of the H$_2$O-abundance. We also note  that the H$_2$O line profile predictions are quite sensitive to the exact value of the kinetic temperature and the dust radiation field (and wavelength-dependent absorption efficiencies) \citep{Decin2010}, which in part might also explain the difference between the value we derive [ortho-H$_2^{16}$O/H$_2$]=$5\times10^{-5}$
and the value of \citet{Maercker2008A&A...479..779M} [ortho-H$_2^{16}$O/H$_2$]=$3.5\times10^{-4}$. A comparison with the ISO-LWS data will be presented in Decin et al. (\emph{in prep.}).

For the oxygen-rich Mira W Hya, \citet{Zubko2000ApJ...544L.137Z} derived an OPR value of 1:1.3 and \citet{Barlow1996A&A...315L.241B} reported a value of 1:1. However, \citeauthor{Barlow1996A&A...315L.241B} noted that their derived value is quite uncertain, due to the high opacities in the water lines used. \citet{Justtanont2010} derived an OPR of 2.1:1 for the S-type AGB $\chi$ Cyg. Our study here is the first time that the OPR value in an oxygen-rich Mira is determined from  a combination of emission lines of H$_2^{16}$O, H$_2^{17}$O, and H$_2^{18}$O. The advantage of using the rarer isotopologs is that the opacity of the lines is lower, making the line intensities more sensitive to the OPR value. The observed para-water lines are consistent with an OPR value of 3:1 ($\pm0.4$). The lowest energy level of para-H$_2$O is $\sim$34\,K below that of ortho-H$_2$O. When water forms in the gas phase via exothermic reactions the energy released is much greater than this energy difference and the OPR reflects the high-temperature ($\sim$50\,K) thermodynamic 3:1 ratio of the statistical weights between the species. The derived OPR value of 3:1 confirms that water in IK Tau is formed in warm and dense regions of the envelope where the chemistry is in thermodynamical equilibrium. 

\paragraph{Isotope ratios:}
Assuming the same photodissociation radius as for the main isotopolog, the isotopic ratios we derive for IK~Tau are H$_2^{16}$O/H$_2^{17}$O=600 \textbf{($\pm 150$)} and  H$_2^{16}$O/H$_2^{18}$O=200\textbf{($\pm 50$)},  hence well below the solar values   \citep[$^{16}$O/$^{17}$O$\sim$2632 and $^{16}$O/$^{18}$O$\sim$499;][]{Asplund2009ARA&A..47..481A}. Interpreting the derived isotopic ratios in terms of nucleosynthesis and subsequent dredge-ups or extra mixing processes is quite complex \citep[e.g.][]{Harris1985ApJ...292..620H, Harris1987ApJ...316..294H, Karakas2010ApJ...713..374K}. In stars that are sufficiently massive to undergo CNO-cycle hydrogen burning, the low initial $^{17}$O abundance (assumed to be solar) is enhanced. When helium burning begins inside the hydrogen-burning shell, $^{17}$O is expected to be completely destroyed in the region where maximum hydrogen burning occurs. The isotope $^{18}$O, on the other hand, is expected to be  destroyed during hydrogen burning, so that it virtually disappears from the hydrogen-burning zone and from the hydrogen-exhausted CNO equilibrium zone within it. When helium burning starts, the $^{18}$O abundance might slightly increase. A succession of convective mixings  brings to the surface material that is affected by these nuclear transformations. Calculations by \citet{Harris1985ApJ...292..620H} show that in every star that becomes a red giant star (M$\ga$0.8\,\Msun), the initial $^{16}$O/$^{17}$O ratio decreases to $\sim$440 during the first dredge-up, while the $^{16}$O/$^{18}$O slightly increases. The second dredge-up occurs at the end of core helium burning only for the most massive intermediate-mass stars (M$\ga$4.5\,\Msun). The estimated $^{16}$O/$^{17}$O ratio ranges between 150 and 500, while the $^{16}$O/$^{18}$O ratio slightly increases. The third dredge-up occurs in the subsequent helium shell-burning phase for stars $\ge$2\,\Msun,  and is expected to yield isotopic ratios of $^{16}$O/$^{17}$O $\le200$. If hot bottom burning occurs (for stars above 3--4\,\Msun), the $^{16}$O/$^{17}$O ratio will be of the order of 20--50. That the $^{16}$O/$^{17}$O ratio is around 600, implies that the first  but no subsequent dredge-ups occurred and constrains the initial mass of IK Tau to be within 1--2\,\Msun. Alternatively, if the star is more massive than 2 $M_\odot$ and the third dredge-up has occurred (but has not turned the star into a carbon-rich star), transferred material from a post-third dredge-up envelope must have had a $^{16}$O/$^{17}$O ratio $\leq 200$. This implies that in stars with $^{16}$O/$^{17}$O $\sim 600$ the transferred material has been heavily diluted by material from the star's own envelope with much higher $^{16}$O/$^{17}$O ratios. From Fig.~5 of \citet{Harris1987ApJ...316..294H}, it is estimated that IK Tau has a low  s-process neutron exposure, $\tau_0$ $\le 0.1$, implying a low absolute enhancement of the s-process elements and only a few third dredge-up events, consistent with IK Tau still being an oxygen-rich AGB star. 

Hitherto, the lower than solar $^{16}$O/$^{18}$O ratios cannot be explained by any stellar evolution model in the literature. However, IK Tau is not the only Galactic star with a low $^{16}$O/$^{18}$O ratio \citep[see Fig.\ 3 in][]{Karakas2010ApJ...713..374K}; some barium stars analyzed by \citet{Harris1985ApJ...292..620H}  also have a low $^{16}$O/$^{18}$O value. It is anticipated that the observations of other evolved stars in the framework of the HIFISTARS programme (P.I.\ V.\ Bujarrabal) will add new information to this discussion.

\subsection{Thermodynamical structure of the envelope} \label{envelope_structure}
Being a simple diatomic molecule with a well understood energy diagram, CO has been successfully used to study the structure of the CSEs around evolved stars \citep[e.g.,][]{Schoier2002A&A...391..577S, Decin2006A&A...456..549D}. Different transitions can be used to investigate different regions of the envelope, probing the density, the temperature, and the velocity of the CSE. \citet{Decin2010} used the $^{12}$CO J=1--0 to J=7--6 lines to determine the thermophysical structure of the CSE of IK Tau beyond $\sim$100\,\Rstar, based on a non-local thermodynamic equilibrium (non-LTE) radiative transfer analysis of the available transitions. In the first instance, the kinetic temperature and velocity structure of the envelope were calculated by solving the equations of motion of gas and dust and the energy balance simultaneously. To get insight into the structure in the inner wind region, the HCN J=3--2 and J=4--3 transitions were used, since observational evidence exists that HCN is formed close to the star  \citep[$\la$3.85\arcsec,][]{Marvel2005AJ....130..261M}. The Gaussian HCN line profiles  indeed point toward line formation partially in the inner wind where the stellar wind has not yet reached its full terminal velocity \citep{Bujarrabal1991A&A...251..536B}. The results of \citet{Decin2010} infer a wind acceleration that is lower than derived from solving the momentum equation. 

\subsubsection{Temperature structure}
Adopting the thermodynamic structure derived in \citet{Decin2010} (and reproduced in Fig.~\ref{fig:structure_IKTau} of the online Appendix), the theoretical line profiles for the $^{12}$CO J=10--9 and J=16--15 are calculated (see Fig.~\ref{Fig:2}) using the non-LTE radiative transfer code GASTRoNOoM \citep{Decin2006A&A...456..549D, Decin2010}. The $^{12}$CO J=10--9 line is very well reproduced,  while the J=16--15 line exhibits slightly larger deviations. The latter most likely reflects the very difficult calibration of this frequency setting \citep[for which standing waves heavily perturbed the baseline, see][]{Bujarrabal2010}.  This result is consistent with the temperature structure in the region between 20 and 100\,\Rstar\ derived by \citet{Decin2010}.

\subsubsection{Velocity structure}
\begin{figure}
 \includegraphics[height=0.48\textwidth,angle=90]{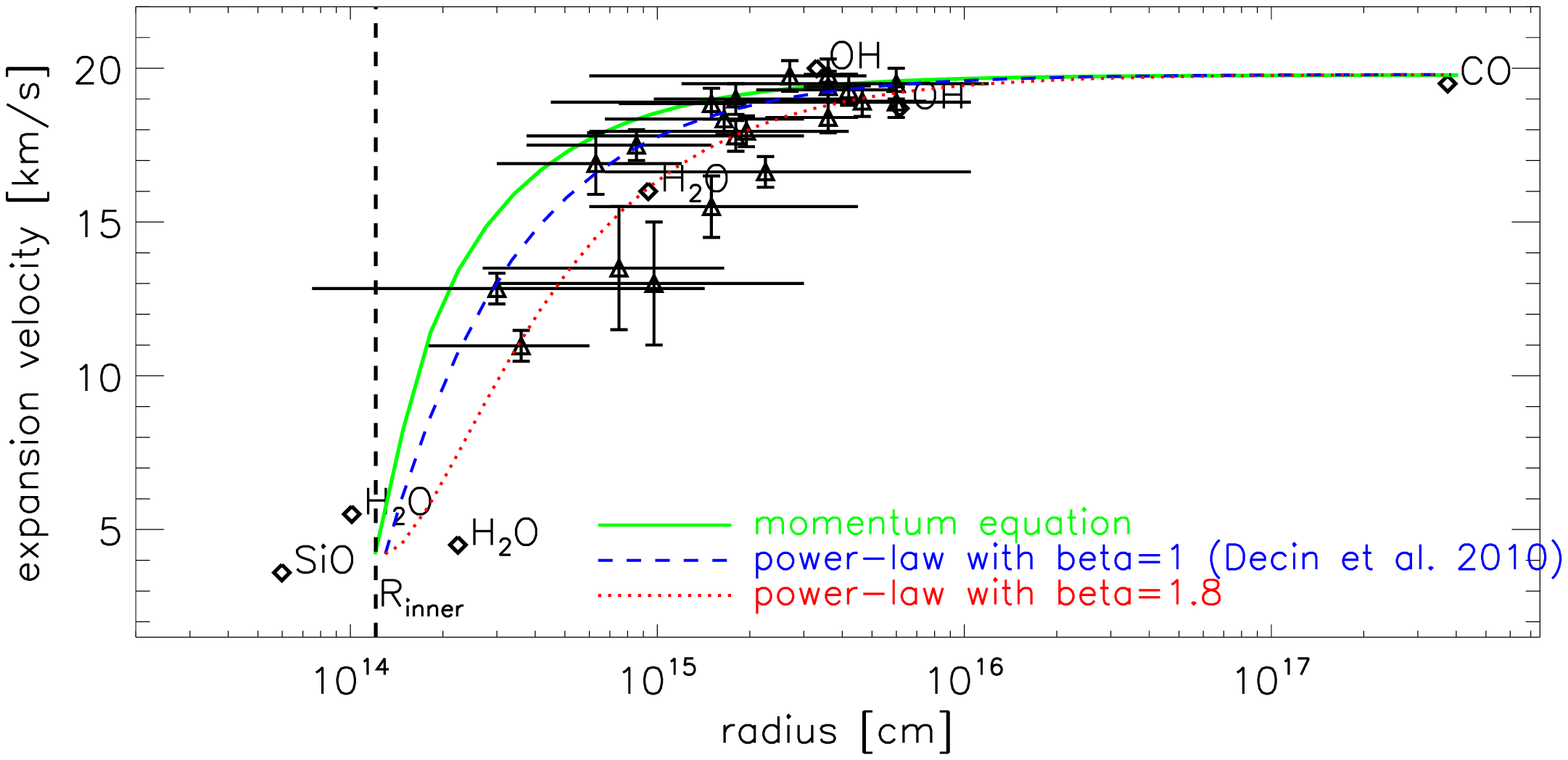}
\caption{Velocity profile of \object{IK~Tau}. Velocity data are obtained from mapping of maser emission: SiO \citep{Boboltz2005ApJ...625..978B}, H$_2$O \citep{Bains2003MNRAS.342....8B}, and OH \citep{Bowers1989ApJ...340..479B}. The CO expansion velocity derived from ground-based CO J=1--0 data is also indicated \citep{Decin2010}. The triangles show the place in the envelope where the line formation is highest for the HIFI data presented in Figs.~\ref{Fig:1}--\ref{Fig:2}, i.e., where $I_{\nu_0}(p)\,p^2$ is at its maximum. The horizontal bars show the minimum and maximum radial distance for the line formation of each individual transition. The vertical bars show the uncertainty on the observed line widths.
The expansion velocity deduced from solving the momentum equation is shown by the full green line. The dashed blue line represents a power law (Eq.~\ref{beta}) with $\beta$\,=\,1 (as used for the modelling  in the online Appendix~\ref{modelling}). For comparison, an even smoother expansion velocity structure with $\beta$\,=\,1.8 is shown with the red dotted line. The vertical dashed black line indicates the dust condensation radius R$_{\rm inner}$.  The velocity at R$_{\rm inner}$ is assumed to be equal to the local sound velocity.}
\label{fig:velocity}
\end{figure}

To constrain the wind acceleration in the CSE, all molecular emission lines as shown in Figs.~\ref{Fig:1}--\ref{Fig:2} were modelled (see  the online Appendix~\ref{modelling}). The line formation region of each molecular line was estimated by considering the range of projected radii  where $I_{\nu_0}(p)\,p\,dp$, with $I_{\nu_0}$ the intensity at the line center and $p$ the impact parameter, exceeds half of its maximum value (see Fig.~\ref{fig:velocity}). 
We note that the radial extent of the line formation region is almost insensitive to the exact value of the velocity structure. Optical depths effects can strongly affect the observed line widths, and detailed radiative transfer modelling (as presented in this Letter) is required to determine the underlying velocity structure. Most of the observed lines have line widths in excess of 17\,km/s. However, a few lines are considerably narrower, and their line formation regions are located in that part of the envelope where the wind has not yet reached its terminal velocity (19\,km/s). Although the line formation regions are quite broad, we find the first observational evidence that the wind acceleration is slower than implied by the momentum equation, corroborating the results of \citet{Decin2010}. Using the classical $\beta$-parametrization \citep[e.g.,][]{Lamers1999isw..book.....L} to simulate the velocity structure, 
\begin{equation}
\varv(r)\simeq \varv_0+(\varv_\infty-\varv_0)\left(1-\frac{R_\star}{r}\right)^\beta\,,
\label{beta}
\end{equation}
where $\varv_0$ is the velocity at the dust condensation radius and $\varv_\infty$ is the terminal velocity, we find that $1\le\beta\le2$. The theoretical line predictions shown in this Letter are computed for $\beta=1$.  The momentum equation, in contrast, matches a much steeper velocity profile with $\beta$=0.6. We note that the velocity structure derived by \citet{Justtanont2010} for $\chi$ Cyg is compliant with a $\beta$-value of 0.9. As discussed by \citet{Decin2010}, there are several possible causes for  a less steep velocity structure. We summarize that a slower wind velocity may be caused by incomplete momentum coupling, dust emission being slightly optically thick to the stellar radiation, the fact that not all dust species are formed at the same time and at the same radial distance, and/or that some dust species are inefficient as wind drivers. 

\section{Conclusions}

Using the HIFI spectrometer onboard Herschel, we have observed the evolved oxygen-rich Mira star, IK Tau, in 31 molecular emission lines.  For the first time, several lines of ortho and para H$_2^{16}$O, and the rarer isotoplogs H$_2^{17}$O and H$_2^{18}$O have been detected. We have deduced a total water content (relative to H$_2$) of $6.6 \times 10^{-5}$, and an ortho-to-para ratio of 3:1. The measured thermal emission from water therefore points toward TE chemistry, as opposed to pulsationally induced non-equilibrium chemistry, grain surface reactions, or evaporation of icy bodies. For the isotopic ratios, we find subsolar values H$_2^{16}$O/H$_2^{17}$O=600 and H$_2^{16}$O/H$_2^{18}$O=200.

The high-excitation lines of $^{12}$CO, $^{13}$CO, $^{28}$SiO, $^{29}$SiO, $^{30}$SiO, HCN, and SO  indicate that the wind acceleration is slower than hitherto anticipated from solving the momentum equation.  These results show the great capability of HIFI to study the complex thermodynamical and chemical envelopes of evolved stars in tremendous detail.

\begin{acknowledgements}
HIFI has been designed and built by a consortium of institutes and
university departments from across Europe, Canada and the United States
under the leadership of SRON Netherlands Institute for Space Research,
Groningen, The Netherlands and with major contributions from Germany,
France and the US.  Consortium members are: Canada: CSA, U.Waterloo;
France: CESR, LAB, LERMA, IRAM; Germany: KOSMA, MPIfR, MPS; Ireland,
NUI Maynooth; Italy: ASI, IFSI-INAF, Osservatorio Astrofisico di
Arcetri- INAF; Netherlands: SRON, TUD; Poland: CAMK, CBK; Spain:
Observatorio Astron\'omico Nacional (IGN), Centro de Astrobiolog\'{\i}a
(CSIC-INTA); Sweden: Chalmers University of Technology - MC2, RSS \&
GARD; Onsala Space Observatory; Swedish National Space Board, Stockholm
University - Stockholm Observatory; Switzerland: ETH Zurich, FHNW; USA:
Caltech, JPL, NHSC. HCSS / HSpot / HIPE is a joint development by the
Herschel Science Ground Segment Consortium, consisting of ESA, the NASA
Herschel Science Center, and the HIFI, PACS and SPIRE consortia.
LD  acknowledges financial support from the Fund for Scientific
  Research - Flanders (FWO). This work has been partially supported by the
Spanish MICINN,  program CONSOLIDER INGENIO 2010,  grant ASTROMOL ( CSD2009-00038). R.Sz.\ and M.Sch.\ acknowledge support from grant N 203
393334 from Polish MNiSW. J.C.\ thanks funding from MICINN, grant AYA2009-07304. This research was performed, in part, through a JPL contract funded by the National Aeronautics and Space Administration.
\end{acknowledgements}
\vspace*{-3ex}
\bibliographystyle{aa}
\bibliography{15069}


\Online
\begin{appendix}

\section{Details about the observation strategy and data reduction}\label{App_observations}
The regular DBS mode was used for bands 1 up to 5, while FastDBS was applied to bands 6 and 7 to achieve higher stability with respect to electrical standing waves. Two orthogonal polarizations were measured simultaneously. The double-sideband \citep[DSB,][]{Helmich2010} observations ensure an instantaneous 8\,GHz and 5.2\,GHz frequency coverage by the wideband spectrometer (WBS) for respectively bands 1 up to 5, and bands 6 and 7. The spectral resolution is 0.5\,MHz. Care was taken in choosing local oscillator (LO) frequencies such that no strong lines from the two sidebands would blend, and that at the same time a maximum number of molecular lines would be covered. 

The data were processed with the standard HIFI pipeline using HIPE, and non-stitched Level-2 data were exported using the HiClass tool available in HIPE. Further processing, i.e. blanking spurious signals, baseline removal, stitching of the spectrometer subbands, and averaging, was performed in CLASS. When the quality of the spectra measured in both horizontal and vertical polarization was good, these were averaged to lower the rms noise. This approach is justified since polarisation is not a concern for the presented molecular-line analysis. In all cases, we assumed a side-band gain ratio of one.

All data presented in this Letter were converted from the antenna-temperature ($T_{A}^*$) scale to the main-beam temperature ($T_{\rm{MB}}$) scale according to $T_{\rm{MB}}=T_{A}^*/\eta_{\rm{MB}}$, with values of the main-beam efficiency $\eta_{\rm{MB}}$ (Table\,\ref{tbl:info}) calculated for the LO frequency $\nu_{\rm{LO}}$, according to 
\begin{equation}
 \label{eq:ruze}
\eta_{\rm{MB}}=\eta_{\rm{B}}\times\exp\left(-\left(\frac{\nu_{\rm{LO}}/\rm{GHz}}{6\times10^{3}}\right)^2\right)\times\eta_{\rm{F}},
\end{equation}
with $\eta_{\rm{B}}$=0.72 and $\eta_{\rm{F}}$=0.96 being the beam efficiency in the 0\,Hz frequency limit and the forward efficiency, respectively. The absolute calibration accuracy ranges from 10\% for the lowest frequency lines up to 30\% for the high frequency ($>$1\,THz) lines.

\begin{table}[htp]
\caption{LO frequencies $\nu_{\rm{LO}}$, main beam efficiencies $\eta_{\rm{MB}}$ according to Eq.\,\ref{eq:ruze}, velocity resolution $\Delta \varv$ (for 0.5\,MHz frequency resolution), and operational days (OD) when observations were carried out.}\label{tbl:info}
\begin{center}
 \begin{tabular}{cccc}
\hline\hline\\[-2ex]
$\nu_{\rm{LO}}$ & $\eta_{\rm{MB}}$ & $\Delta \varv$& OD \\
(GHz) & &(km/s)&\\
\hline\\[-2ex]
  564.476 &0.69 &0.27&292	\\
  614.763 &0.68 &0.24&292	\\
  758.772 &0.68 &0.20&297	\\
  975.087 &0.67 &0.15&293	\\
  995.481 &0.67 &0.15&293	\\
  1102.757&0.67 &0.14&294	\\
  1106.735&0.67 &0.14&294	\\
  1157.495&0.67 &0.13&296	\\
  1200.717&0.66 &0.12&296	\\
  1713.596&0.64 &0.09&298	\\
  1757.417&0.63 &0.09&298	\\
  1838.039&0.63 &0.08&298	\\
  1864.539&0.63 &0.08&298	\\
\hline
 \end{tabular}
\end{center}
\end{table}

\begin{table}[htp]
\caption{Observation summary. First column gives the molecular name, second column the transition, third column the frequency of the line, fourth column the integrated intensity, fifth column the noise, and last columns some comments (if necessary). One line is still unidentified, and is indicated with a 'U' in the first column.}
\label{Table:2}
\scriptsize{
\setlength{\tabcolsep}{1mm}
\begin{center}
 \begin{tabular}{clcccl}
\hline
\hline
Molecule & Transition & Freq. [\textbf{GHz}] & I [K km/s] & noise [K] & comments\\
\hline
o-H$_2^{16}$O & $1_{1,0}-1_{0,1}$ & 556.936 & 10.22 & 0.005 &   \\
o-H$_2^{16}$O & $5_{3,2}-4_{4,1}$ & 620.701 & 14.89 & 0.005 & maser \\
o-H$_2^{16}$O & $3_{1,2}-3_{0,3}$ & 1097.365 & 14.44 & 0.036 &   \\
o-H$_2^{16}$O & $3_{1,2}-2_{2,1}$ & 1153.127 & 22.60 & 0.062 &   \\
o-H$_2^{16}$O & $3_{0,3}-2_{1,2}$ & 1716.770 & 30.89 & 0.104 &   \\
o-H$_2^{16}$O & $5_{3,2}-5_{2,3}$ & 1867.749 & 3.45 & 0.118 & standing waves \\
o-H$_2^{16}$O & $3_{2,1}-3_{1,2}$ & 1162.912 & 13.70 & 0.062 &   \\
p-H$_2^{16}$O & $2_{1,1}-2_{0,2}$ & 752.033 & 13.71 & 0.015 &   \\
p-H$_2^{16}$O & $5_{2,4}-4_{3,1}$ & 970.315 & 19.34 & 0.026 & maser \\
p-H$_2^{16}$O & $2_{0,2}-1_{1,1}$ & 987.927 & 27.85 & 0.030 &   \\
p-H$_2^{16}$O & $1_{1,1}-0_{0,0}$ & 1113.343 & 25.10 & 0.028 &   \\
p-H$_2^{16}$O & $1_{1,1}-0_{0,0}$ $\nu_2$=1 & 1205.798 & 0.59 & 0.070 & 
tentative, see Fig. 1 \\
p-H$_2^{16}$O & $4_2{,2}-4_{1,3}$ & 1207.639 & 8.21 & 0.070 & maser \\
p-H$_2^{16}$O & $5_{3,3}-6_{0,6}$ & 1716.953 & 1.03 & 0.104 & standing waves \\
o-H$_2^{17}$O & $3_{1,2}-3_{0,3}$ & 1096.414 & 1.20 & 0.036 &   \\
p-H$_2^{17}$O & $1_{1,1}-0_{0,0}$ & 1107.167 & 1.50 & 0.036 &   \\
o-H$_2^{18}$O & $3_{1,2}-3_{0,3}$ & 1095.627 & 2.52 & 0.036 &   \\
p-H$_2^{18}$O & $1_{1,1}-0_{0,0}$ & 1101.699 & 3.57 & 0.028 &   \\
$^{12}$C$^{16}$O & $10-9$ & 1151.985 & 9.61 & 0.062 &   \\
$^{12}$C$^{16}$O  & 16-15 & 1841.35 &2.48 &0.03 & standing waves \\
$^{13}$C$^{16}$O & $9-18$ & 991.329 & 2.43 & 0.030 &   \\
$^{13}$C$^{16}$O & $10-9$ & 1101.350 & 2.75 & 0.028 &   \\
$^{28}$Si$^{16}$O & $14-13$ & 607.599 & 2.31 & 0.005 &   \\
$^{29}$Si$^{16}$O & $13-12$ & 557.179 & 0.89 & 0.005 &   \\
$^{29}$Si$^{16}$O & $26-25$ & 1112.833 & 0.74 & 0.028 & low S/N, see Fig. 2 \\
$^{30}$Si$^{16}$O & $26-25$ & 1099.708 & 0.39 & 0.028 & 
tentative, see Fig. 2 \\
HCN & $7-6$ & 620.304 & 2.65 & 0.005 &   \\
HCN & $13-12$ & 1151.452 & 2.22 & 0.062 & tentative, see Fig. 2 \\
SO & $13_{14}-12_{13}$ & 560.178 & 0.20 & 0.005 &   \\
NH$_3$ & $1_0-0_1$ & 572.498 & 4.93 & 0.005 &   \\
U &   & 988.255 & 1.37 & 0.030 & or at 1003.215 MHz \\
\hline
\end{tabular}
\end{center}}
\end{table}

\section{Radiative transfer modelling} \label{modelling}

\begin{figure}[htp]
\begin{center}
\includegraphics[width=.48\textwidth,angle=0]{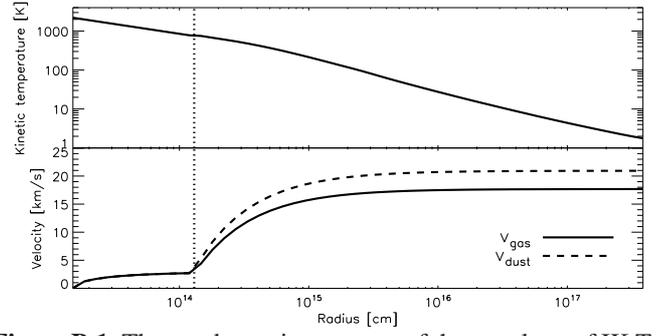}
\vspace*{2ex}
\caption{Thermodynamic structure of the envelope of \object{IK Tau} 
  derived from the $^{12}$CO J=1--0 to J=7--6 and HCN J=3--2 and J=4--3 rotational line transitions for the stellar parameters given  in Table~\ref{Table:1} \citep{Decin2010}. The start of the dusty
  envelope, R$_{\rm{inner}}$, is indicated by the dotted line. }
\label{fig:structure_IKTau}
\end{center}
\end{figure}

The molecular emission lines shown  in Figs.~\ref{Fig:1}--\ref{Fig:2}, were modelled using the non-LTE radiative transfer code GASTRoNOoM \citep{Decin2006A&A...456..549D}. Last updates to the code and a discussion of the available line lists and collisional rates can be found in \citet{Decin2010}. The thermodynamical structure (see Fig.~\ref{fig:structure_IKTau}) determined by  \citet{Decin2010} was confirmed using the new HIFI observations (Sect.~\ref{envelope_structure}). Modelling the CO and H$_2$O lines provides insight into to the cooling/heating rates by transitions of these molecules \citep[see also][]{Decin2006A&A...456..549D, Decin2010}. As can be seen in Fig.~\ref{fig:heat_cool},  H$_2$O transitions provide the main cooling agent in the region up to $\sim$2$\times10^{15}$\,cm, while adiabatic cooling takes over for the region beyond $\sim$2$\times10^{15}$\,cm.

\begin{figure}[htp]
\begin{center}
\includegraphics[height=.48\textwidth,angle=90]{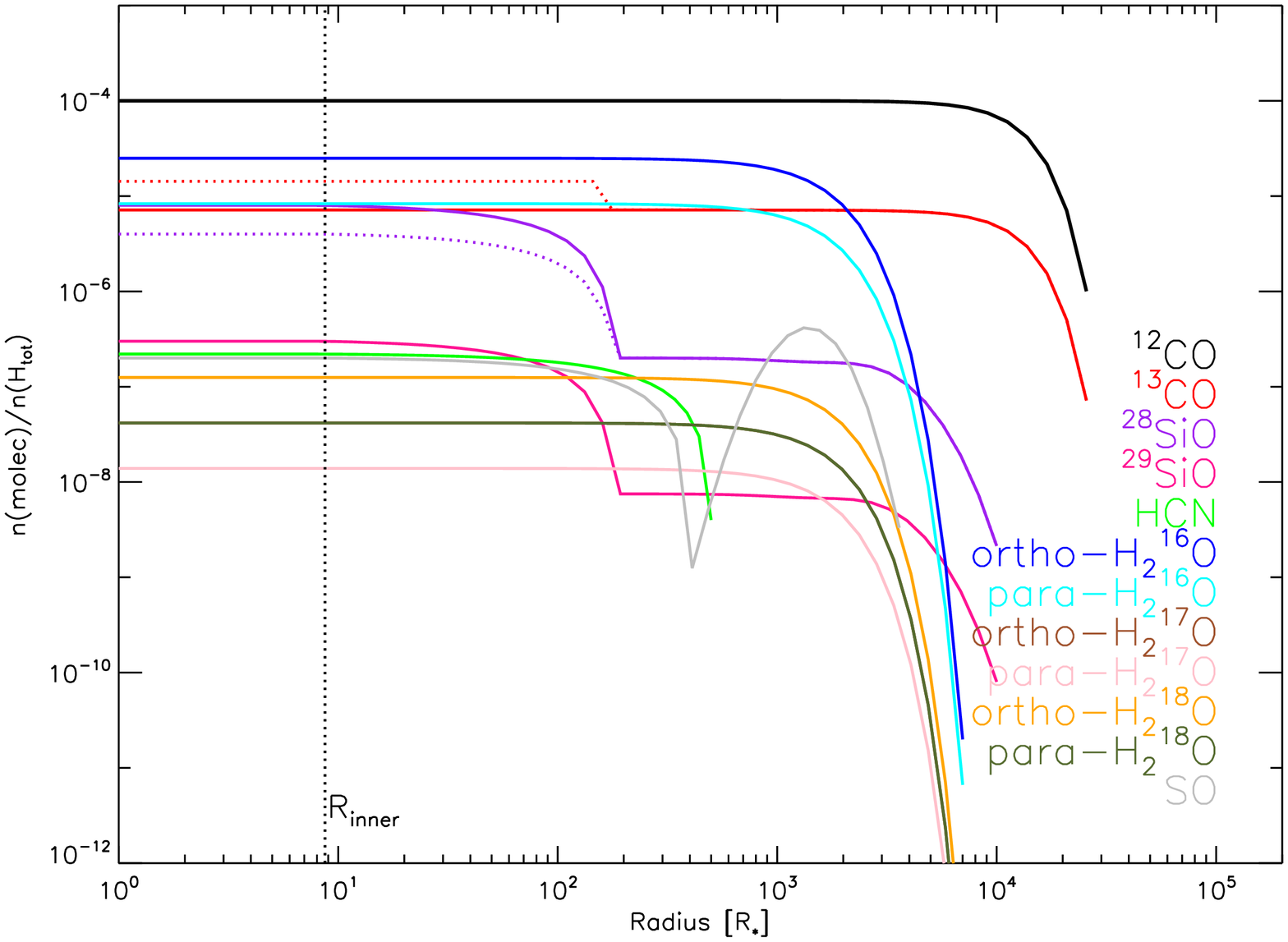}
\vspace*{2ex}
\caption{Fractional abundance stratifications for $^{12}$CO, $^{13}$CO, $^{28}$SiO,  $^{29}$SiO, HCN, SO, ortho-H$_2^{16}$O, para-H$_2^{16}$O, ortho-H$_2^{17}$O, para-H$_2^{17}$O, ortho-H$_2^{18}$O, and para-H$_2^{18}$O. For all molecules (except water), the full line represents the results  obtained by \citet{Decin2010}. For  $^{13}$CO and $^{28}$SiO, the new results based on the HIFI data are shown in dotted lines. The fractional abundances for all water isotopologs and isomers are based on the HIFI data presented in this Letter.}
\label{fig:abundances}
\end{center}
\end{figure}

\begin{figure}[htp]
\begin{center}
\includegraphics[width=.48\textwidth,angle=0]{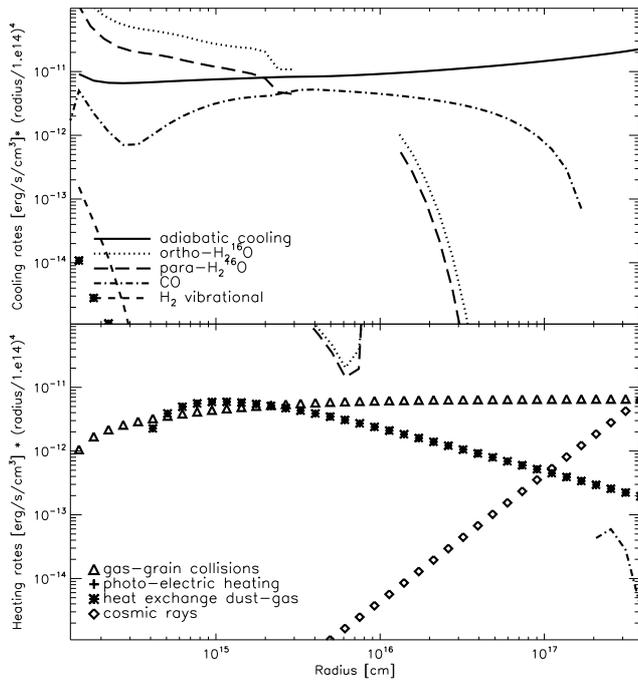}
\vspace*{2ex}
\caption{Cooling and heating rates in the envelope of \object{IK Tau} due to different processes \citep[for details, see][]{Decin2006A&A...456..549D}. As can be seen in the plot, both CO and H$_2$O transitions mainly cool the envelope, but in certain restricted ranges can heat the envelope. }
\label{fig:heat_cool}
\end{center}
\end{figure}

In the first instance, the molecular abundance stratifications derived by \citet{Decin2010} were assumed to model the $^{13}$CO, $^{28}$SiO, $^{29}$SiO, $^{30}$SiO, HCN, and SO lines (see full black lines in Fig.~\ref{Fig:2} and full lines in Fig.~\ref{fig:abundances}). Using the new HIFI observations, the $^{13}$CO and $^{28}$SiO abundance fractions were refined in the inner envelope (see dashed black lines in Fig.~\ref{Fig:2} and dashed lines in Fig.~\ref{fig:abundances}). 

The radiative transfer modelling for water included the 45 lowest levels of the ground state and first vibrational state (i.e.\ the bending mode $\nu_2\,=\,1$ at 6.3\,$\mu$m) for all isotopologs. Level energies, frequencies, and Einstein A coefficients were extracted from the HITRAN water line list \citep{Rothman2009JQSRT.110..533R}. The H$_2$O-H$_2$ collisional rates were taken from \cite{Faure2007A&A...472.1029F}. The effect of including excitation to the first excited vibrational state of the asymmetric stretching mode ($\nu_3=1$) was tested, and was found to be negligible \citep{Decin2010}. 

A good agreement was found for the HCN(7--6) line proving that the inner abundance fraction [HCN/H$_{\rm{tot}}$] is $\sim2.2 \times10^{-7}$. The $^{13}$CO J=9--8 and J=10--9 lines are somewhat underpredicted assuming a $^{12}$CO/$^{13}$CO ratio of 14 as obtained by \citet{Decin2010}. Decreasing this ratio to 7 yields a better fit,  but we point out that  the line profiles are quite noisy. The $^{13}$CO fractional abundance was obtained assuming the same photodissociation radius as for $^{12}$CO \citep{Mamon1988ApJ...328..797M}. However, if the effect of less self-shielding of $^{13}$CO (compared to $^{12}$CO) were more important than estimated by \citet{Mamon1988ApJ...328..797M}, the photodissocation radius of $^{13}$CO would be smaller, affecting the low excitation rotational transitions more than the higher excitation lines observed by HIFI. Another possibility might be that the velocity structure is steeper than the $\beta=1$ power law now assumed in the region between $\sim$20 and 150\,\Rstar, where these high-excitation lines are mainly formed. Since a constant mass-loss rate is assumed, this would imply a lower density in this region, and hence a higher $^{13}$CO abundance fraction to produce the correct line intensity. 

Only one higher-excitation $^{28}$SiO line has been observed so far. The J=14--13 transition indicates that the inner wind abundance might be a factor 2 lower than deduced by \citet{Decin2010}, yielding an inner wind abundance of $4 \times 10^{-6}$, decreasing to $2 \times 10^{-7}$ around 180\,\Rstar. The isotopolog line of  $^{29}$SiO(13--12) is very well predicted for an inner abundace of $3 \times 10^{-7}$; the higher excitation J=26--25 line of both $^{29}$SiO and $^{30}$SiO are consistent with the HIFI observations. This implies an isotopic ratio of $^{28}$SiO/$^{29}$SiO of 13. 

Using the abundance pattern determined by \citet{Decin2010}, the SO($13_{14}-12_{13}$) is quite well predicted.

\begin{table}
\caption{(Circum)stellar parameters for IK Tau \citep{Decin2010}. \Teff\ is the effective stellar temperature, \Rstar\ denotes the stellar radius,  \Mdot\ the gas mass-loss rate,  $R_{\rm{inner}}$ the dust condensation radius, $\varv_{\infty}$ the terminal velocity of the wind, and $\varv_{\rm{turb}}$ the turbulent velocity in the wind. The molecular fractional abundances are given relative to H$_{\rm{tot}}$=n(H)+2n(H$_2$), and denote the abundance at the dust condensation radius (see Fig.~\ref{fig:abundances}). Values between parentheses denote refined inner wind abundance values obtained from the new HIFI observations.The fractional abundances for all water isotopologs and isomers are based on the HIFI data presented in this Letter.}
\label{Table:1}
\begin{center}
\vspace*{-1.5ex}
\begin{tabular}{lc|lc}
\hline \hline
\rule[0mm]{0mm}{5mm}{\Teff} [K] & 2200 & $[$CO/H$_{\rm{tot}}]$ & $1 \times 10^{-4}$ \\
$R_{\star}$ [$10^{13}$\,cm] & 1.5 &   $^{12}$CO/$^{13}$CO & 14 (7) \\
 {\Mdot} [{\Msun}/yr] & $8 \times 10^{-6}$ &  $[^{28}$SiO/H$_{\rm{tot}}]$ & $8 \times 10^{-6}$ ($4 \times 10^{-6}$) \\
distance  [pc] & 265 &  $[^{29}$SiO/H$_{\rm{tot}}]$ & $3 \times 10^{-7}$ \\
$R_{\rm{inner}}$ [\Rstar] & 8.7&  $[^{30}$SiO/H$_{\rm{tot}}]$ & $1 \times 10^{-7}$ \\
$\varv_{\infty}$    [km\,s$^{-1}$] &  17.7 &    $[$HCN/H$_{\rm{tot}}]$ & $2.2 \times 10^{-7}$\\
$\varv_{\rm{turb}}$     [km\,s$^{-1}$] &  1.5 & [SO/H$_{\rm{tot}}]$ & $1 \times 10^{-6}$ ($1 \times 10^{-5}$)\\
\hline
\rule[0mm]{0mm}{5mm}$[$ortho-H$_2^{16}$O/H$_{\rm{tot}}]$ & $2.5 \times 10^{-5}$ & OPR & 3:1\\
H$_2^{16}$O/H$_2^{17}$O & 600 & H$_2^{16}$O/H$_2^{18}$O & 200 \\
\hline
\end{tabular}
\end{center}
\end{table}

\end{appendix}

\end{document}